\documentclass[aps,prc,showpacs,twocolumn]{revtex4}
\usepackage{graphicx}
\usepackage{dcolumn}
\usepackage{bm}
\usepackage{longtable}
\usepackage{hyperref}
\usepackage{natbib}
\usepackage{amsmath}
\begin{document}
\title{$\beta$-decay properties for neutron-rich Kr-Tc isotopes from deformed
pn-QRPA calculations with realistic forces}
\author{Dong-Liang Fang$^{a,b}$, B. Alex Brown$^{a,b,c}$ and Toshio
Suzuki$^{d,e}$}
\affiliation{$^a$National Superconducting Cyclotron Laboratory, Michigan State
University, East Lansing, Michigan 48824, USA}
\affiliation{$^b$Joint Institute for Nuclear and Astrophysics, Michigan State
University}
\affiliation{$^c$Department of Physics and Astronomy, Michigan State University,
East Lansing, MI 48824, USA}
\affiliation{$^d$Department of Physics, College of Humanities and Sciences,
Nihon University, Sakurajosui 3-25-40, Setagaya-ku, Tokyo 156-8550, Japan}
\affiliation{$^e$National Astronomical Observatory of Japan,
Mitaka, Tokyo 181-8588, Japan}
\begin{abstract}
In this work we studied $\beta$-decay properties for deformed neutron-rich
nuclei in the region Z=36-43. We use the
deformed pn-QRPA methods with the realistic CD-Bonn forces, and include both the
Gamow-Teller and first-forbidden types of decays in the calculation.
The obtained $\beta$-decay half-lives and neutron-emission probabilities
of deformed isotopes are compared with experiment as well as 
with previous calculations.
The advantages and disadvantages of the method are discussed.
\end{abstract}
\maketitle
\section{introduction}
Decay properties such as the half-lives and $\beta$-delayed neutron emission
probabilities are important inputs for the simulations of the r-process
nucleosynthesis which is believed to be responsible for the production of heavy
elements in our universe.
In order to understand the observed elements abundance,
one needs measurements together with models for making
accurate predictions for these global properties of atomic
nuclei out to the neutron drip-lines, especially those neutron-rich nuclei
along r-process paths \cite{MPK03,bab2}.

Recently, a group from RIKEN has performed a series of half-life measurements
for
the neutron-rich Kr-Tc isotopes \cite{RIKEN11}. For some of these nuclei, some
differences from the previous measurement has been found, while for others, the
half-lives are measured for the first time. These measurements give us more
information for exotic neutron-rich nuclei and also offer us more information
relevant for the r-process flow path around the A=130 peak.

These new measurements serve as good tests or
constraints for theories. The theory for such calculations from gross to
microscopic have been developed for decades. There have
been global estimations of half-lives
from the gross theories such as those in Ref. \cite{TY69} which treats the
half-lives
as functions of the Q-values, proton (Z) and neutron (N) numbers.
More microscopic
methods have been developed in Ref. \cite{MPK03} with deformed pn-QRPA
methods for
the Gamow-Teller (GT) type decays with the residual interaction from the
phenomenological pn forces in $J^\pi=1^+$ channel. Because of the
phenomenological nature of the forces designed only for the GT channel,
the First-Forbidden (FF) part was
estimated
by gross theory instead of direct calculations \cite{MPK03}.
Despite these approximations,
Ref. \cite{MPK03} gives a reasonable average agreement with half-lives over the
nuclear
chart and for the new RIKEN data, but there are order of magnitude deviations
between experiment and theory in many cases.

As an improvement, we proposed the adaption of realistic
forces for the residual interactions. This method was first developed by the
T\"ubingen group \cite{YRFS08,FFRS11} for $\beta\beta$-decay. The advantage
of this method is that we are not restricted to GT channels. With
just two parameters for the residual interactions, we can calculate all
possible decay channels such as those transitions from
$0^+$ ground states to $0^-,1^-,2^-$ final
states.
This gives explicit results from
microscopic calculations for FF decays which in some nuclei may play an important
role. On the other hand, inclusion of all spin-parities can give us better
determinations of the spectra of odd-odd nuclei that makes the
decay energy required for phase-space factors much more accurate. Both are advantages
which can make a better prediction for $\beta$-decay properties.

This article is arranged as follows; in Sec. II we introduce the
formulations of the calculations for $\beta$-decay properties and the
many-body approach we adopt, in Sec. III the results and comparison
to experiment and other models are discussed. The Conclusions are
given in Sec. IV.

\section{formalism}
The half-lives of $\beta$-decaying isotopes can be expressed as:
\begin{eqnarray}
t_{1/2}=\ln 2/\Gamma,
\end{eqnarray}
where $\Gamma$ is the decay width and has the form:
\begin{eqnarray}
\Gamma=\sum_{j}\Gamma_j=\sum_{j} \frac{1}{2\pi^3}\frac{m_e^5
c^4}{\hbar^7}f_j(Z,R,\epsilon_0)|M_j|^2.
\label{thf}
\end{eqnarray}
Here the sum runs over all the possible states $j$ for the final nuclei, $f_j$
is the phase-space factor for state $j$, which is a function of the nucleus
radius
$R$, nuclear charge $Z$ and the energies of the emitted electron
$\epsilon_0=E_e/m_e=(Q-E_j)/m_e$ in unit of the electron mass $m_e$, while $M_j$
is the nuclear matrix element. In this work, we consider two kinds of decays,
the Gamow-Teller (GT) and first-forbidden (FF) decays.
The phase-space factors for both decays are expressed
analytically in Ref. \cite{SYKO11} and references therein. The nuclear matrix
elements describe the nuclear-structure part of the $\beta$-decay. They can be
expressed in the form:
\begin{eqnarray}
M_{j}^{I}=\langle j|O^{I}|i\rangle,
\end{eqnarray}
where $|i\rangle$ is the ground state for the parent nucleus and $O^{I}$ are the
transition operators for $\beta$-decay and have different
forms and selection rules for different types of decay $I$. From the expression
in
\eqref{thf} for the half-lives for decays to each final state, one
needs the information for the excitation energies of the final states and the
matrix element between the initial and final states. For different type of
nuclei, we will use different many-body treatments as explained below.

Various methods have been applied to this
calculation each with some limitations. The large-basis shell model
can account for many of the correlations, but the number of
orbitals that can be considered is restricted by the computational limitations
to dimensions of on the order of 10$^{10}$, and in practice is used
only for the Gamow-Teller decay in light nuclei ($A \leq 60$). All of the
nuclei considered here are outside of this range.

Thus one needs
other methods that involve various approximations.
In our case, we adopt the deformed
version of pn-QRPA with realistic forces as first introduced in Ref.
\cite{YRFS08}.
With the adiabatic Bohr-Mottelson approximation, one can prove the equivalence
of the calculations performed in the laboratory systems and intrinsic systems
\cite{MR90} (In the intrinsic system, the $z$ axis is attached to the symmetric
axis of the nucleus). Thus we will perform our calculation in the intrinsic
system without the consideration of rotations of the nucleus, and adopt the
axially symmetric wave functions in the intrinsic frame. Details of the
wave function calculation is described in Ref. \cite{YRFS08} as well as the
treatment
of BCS
pairing in the deformed nuclei. With the BCS pairing, one defines the
Boglyubov quasi-particle creation and annihilation operators:
\begin{eqnarray}
\left(
\begin{array}{c}
\alpha^\dagger_\tau \\
\tilde{\alpha}_\tau
\end{array}
\right)=
\left(
\begin{array}{c c}
u_\tau & v_\tau \\
-v_\tau&u_\tau
\end{array}
\right)
\left(
\begin{array}{c}
c^\dagger_\tau \\
\tilde{c}_\tau
\end{array}
\right).
\end{eqnarray}
where the annihilation operators annihilate the BCS
vacuum, $\alpha_\tau |BCS\rangle=0$. 
Here $c$ and $c^\dagger$ are single-particle annihilation and creation
operators, and $u$'s and $v$'s are BCS coefficients from the solutions of BCS
equation.

The intrinsic excited states $|K^\pi,m\rangle$ are then generated by the QRPA
creation operators acting on the ground states \cite{FFRS11}:
\begin{eqnarray}
|K^\pi,m\rangle&=&Q^\dagger_{K^\pi,m} |0^+_{g.s.}\rangle \nonumber \\
Q^\dagger_{K^\pi,m}& =&\sum_{p,n} X^m_{pn,K^\pi}
A^\dagger_{pn,K^\pi}-Y^m_{pn,K^\pi} \tilde{A}_{pn,K^\pi}.
\end{eqnarray}
Here the two quasi-particle creation and annihilation operators are defined as:
$A^\dagger_{pn,K^\pi}=\alpha_p^\dagger \alpha_{\tilde{n}}^\dagger$ and
$\tilde{A}_{pn,K^\pi}=\alpha_{\tilde{p}} \alpha_n$, with the selection rule:
$K=\Omega_p-\Omega_n$. In order to obtain the forward and backward amplitudes
X's and Y's, one needs to solve the QRPA equations:
\begin{eqnarray}
\left(
\begin{array}{c c}
A(K^\pi)&B(K^\pi)\\
B(K^\pi)&A(K^\pi)
\end{array}
\right)
\left(
\begin{array}{c}
X^m_{K^\pi}\\
-Y^m_{K^\pi}
\end{array}
\right)=\omega_{K^\pi,m}
\left(
\begin{array}{c}
X^m_{K^\pi}\\
-Y^m_{K^\pi}
\end{array}
\right).
\end{eqnarray}
The expression of matrices $A$ and $B$ and the details of the interactions are
discussed in Refs. \cite{YRFS08,FFRS11}. The QRPA equations can be solved by
diagonaliztion following the method in Ref. \cite{RS80}. The solutions contain
the
information of the energies from the eigenvalues and the structure information
from the forward and backward amplitudes. With the states constructed from the
solutions of QRPA equations, we can calculate the beta-decay half-lives. Here we
briefly introduce the details of the calculations for different types of nuclei.

First, for the decays of even-even nuclei
[the first even (odd) refers to the proton number Z and second
even (odd) refers to the neutron number N], the parent nuclei have ground states
with all the neutrons and protons paired. Thus it is the BCS vacuum with
$J^\pi=0^+$. The excited states for daughter nuclei in pn-QRPA formalism are
just those we constructed above, we choose the states with the lowest
eigenvalues to be the ground states of the corresponding odd-odd nuclei which
are the decay products. (This requires a calculation over all the possible spin
projections and parities $K^\pi$). The excitation energies for each
states are $E_{K^\pi,m}=\omega_{K^\pi,m}-\omega_{g.s.}$. The matrix elements of
the decay to the $m$th states with spin-parity $K^\pi$ can then be derived as
following:
\begin{eqnarray}
M^{I}_{K^\pi,m}=\langle K^\pi,m|O^I_{M}|0^+_{g.s.}\rangle \nonumber\\
=\sum_{pn}\delta_{KM} \langle p|O^I_{M}|n\rangle (X^m_{pn,K^\pi}u_p
v_n+Y^m_{pn,K^\pi}v_p u_n).
\end{eqnarray}
Here $\langle p|O^I_{M}|n\rangle$ is the single-particle transition matrix
element in
the deformed basis, which can be written as a decomposition over the reduced
matrix elements in spherical harmonic oscillator basis \cite{YRFS08}: $\langle
p|O^I_{M}|n\rangle=\sum_{\eta_p,\eta_n}F^{IM}_{p\eta_p,n\eta_n}\langle
\eta_p||O^I||\eta_n\rangle/\sqrt{2I+1}$. Here for GT decay the operator has the
form $O^{GT}_K=\sigma_K\tau^+$, with the selection rules $\Delta K=0,\pm 1$ and
$\Delta \pi=1$, while the expressions for the first-forbidden beta decay
are more complicated with six components with different spin-parity as
introduced
in Ref. \cite{WM61}. We will not give the explicit expression for these
components
here, but mention that of which two have the selection rules $\Delta K=0$,
$\Delta\pi=-1$, three have the selection rules $\Delta K=0,\pm 1$,
$\Delta\pi=-1$ and one has $\Delta K=0,\pm 1,\pm 2$, $\Delta\pi=-1$.

For odd-mass nuclei, we follow the method in Ref. \cite{MPK03},  except that
the $\Delta v=0$ case
(defined in Ref. \cite{MR90}), we consider only the single-particle transitions
without the corrections from particle-vibration couplings for simplicity. The
states of the odd-mass nuclei can be described as one corresponding
quasi-particle (proton or neutron) excitation on the even-even ground states:
\begin{eqnarray}
|Z\pm1,N,i\rangle=\alpha_{p,i}^\dagger |Z,N,0^+\rangle \nonumber \\
|Z,N\pm1,i\rangle=\alpha_{n,i}^\dagger |Z,N,0^+\rangle,
\label{sings}
\end{eqnarray}
or a pn-QRPA excitation state with one spectator single particle (or hole):
\begin{eqnarray}
|Z+1,N,im\rangle=\alpha_{n,i}^\dagger |K^\pi;m\rangle=\alpha_{n,i}^\dagger
Q^\dagger_{K^\pi,m}|Z,N,0^+\rangle \nonumber \\
|Z,N-1,im\rangle=\alpha_{p,i}^\dagger |K^\pi;m\rangle=\alpha_{p,i}^\dagger
Q^\dagger_{K^\pi,m}|Z,N,0^+\rangle.
\label{colls}
\end{eqnarray}
Here the label $i$ is the spectator nucleon which doesn't participate in the
decay process.
The ground states of these nuclei are the one quasi-particle states with the
lowest quasi-particle energies $E_{\tau,0}$; here $\tau$ can be proton or
neutron. Energies for the states in \eqref{sings} are simply the differences
between the quasi-particle energies $E=E_i-E_0$. While for the states in \eqref{colls},
we use different treatments for the energies as that in Ref. \cite{MPK03}:
$E=Q-Q_{EE}+E_m$, $E_m$ is the actual pn-QRPA excitation energy in the even-even
system and $Q-Q_{EE}$ accounts for the difference of the Q values between the
odd mass isotope and the corresponding even-even one. The matrix elements for
the $\Delta v=0$  single-particle transitions are
then expressed as the leading order terms \cite{MR90}:
\begin{eqnarray}
\langle Z+1,N,i|O^I_{M}|Z,N+1,0\rangle=v_{p,i}v_{n,0}\langle
p,i|O^I_{M}|n,0\rangle\nonumber \\
\langle Z,N-1,i|O^I_{M}|Z-1,N,0\rangle=u_{p,i}u_{n,0}\langle
p,i|O^I_{M}|n,0\rangle,
\label{oddsin}
\end{eqnarray}
and for the $\Delta v=1$ spectator case (also defined in Ref. \cite{MR90}):
\begin{eqnarray}
\langle Z+1,N,0m|O^I_{M}|Z,N+1,0\rangle \nonumber \\
=\sum_{p,n\ne
n_0}\delta_{KM}(X_{pn,K^\pi}^mu_{p}v_{n}+Y_{pn,K^\pi}^mv_{p}u_{n})\langle
p|O^I_{M}|n\rangle\nonumber \\
\langle Z,N-1,0m|O^I_{M}|Z-1,N,0\rangle \nonumber \\
=\sum_{p\ne
p_0,n}\delta_{KM}(X_{pn,K^\pi}^mu_{p}v_{n}+Y_{pn,K^\pi}^mv_{p}u_{n})\langle
p|O^I_{M}|n\rangle.
\label{oddcol}
\end{eqnarray}
Here contributions from the orbit occupied by the spectator particle (hole) are
excluded.

Finally, we discuss the case of odd-odd nuclei following the
treatment from Fig. 3 in Ref. \cite{MR90}. In this scenario, the
collective effect is excluded, and for the ground states of the odd-odd
nuclei, one has simply one-neutron particle and one-proton hole acting on the
even-even ground states. 
For the daughter even-even nuclei, the ground states are obviously the
BCS vacuum, while for the excited states, there exists two different
types: the first case is the two-quasiparticle excitation neglecting the
collectivity, the
excitation energies are: $E=E_{p,i}$ + $E_{p,0}$ 
(lower left panel in Fig. 3 of
Ref. \cite{MR90}) or $E=E_{n,i}$ + $E_{n,0}$
(upper right panel in Fig. 3 of Ref. \cite{MR90});
the second case is the pn-QRPA excitation acting on the odd-odd ground, the
energies
are $E=Q-Q_{EE}+E_m$ (upper left panel in Fig. 3 of Ref. \cite{MR90}), $Q_{EE}$
and
$E_m$ have the same meanings as in the odd mass case. In the first case, the
transition matrix elements are just those of \eqref{oddsin} with one spectator
proton hole or neutron particle. The special case is when the final states is
the ground states with all neutrons and protons paired, in this case, the
excitation energies are $0$ instead of $2E_0$(lower right panel in Fig. 3 of
Ref. \cite{MR90}).

The same matrix elements as in \eqref{oddcol} are adopted with
the exclusion of contributions from the orbits occupied by the unpaired neutron
(particle) and proton (hole) of the odd-odd ground states.
In this approximation, the result is independent of the odd-odd ground state
spin, but
the lack of collectivity will give an
over-estimation on the excitation energy of the two quasi-particle final states.
This makes the beta-decay Q values too low, and increases the beta lifetime.
However, as discussed further below,
the $\beta$-decay lifetimes for odd-odd nuclei are not so important
for the r-process path.

Following the definition in Ref. \cite{MPK03}, the overall measure of error in
the deviation of theory from experiment is defined by:
\begin{eqnarray}
r=\log_{10}(\frac{t_{1/2,calc}}{t_{1/2,exp}}) \nonumber \\
\Sigma_{r}=[\frac{1}{n}\sum_{i=1}^{n}(r^i)^2]^{1/2} \nonumber \\
\Sigma_{r}^{10}=10^{\Sigma_{r}}
\label{err}
\end{eqnarray}
Here, $\Sigma_{r}^{10}$ is defined to be the total "error". This will be used later
for quantitative estimation on the quality of calculations.

\begin{figure}
\includegraphics[scale=0.45]{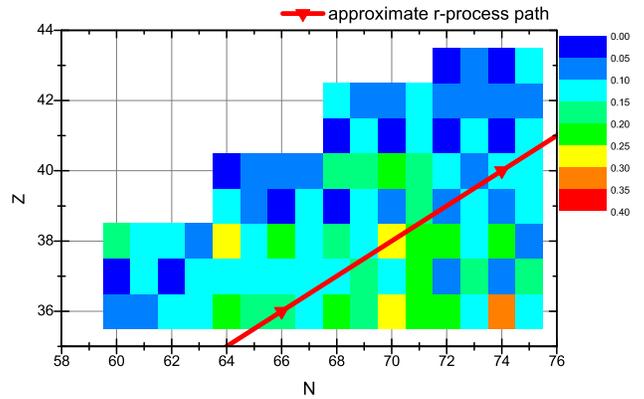}
\caption{Illustrations of calculated branching ratios of the first-forbidden
decay
for different isotopes. The diagonal line indicates the approximate r-process
path.}
\label{ff}
\end{figure}

\begin{figure}
\includegraphics[scale=0.45]{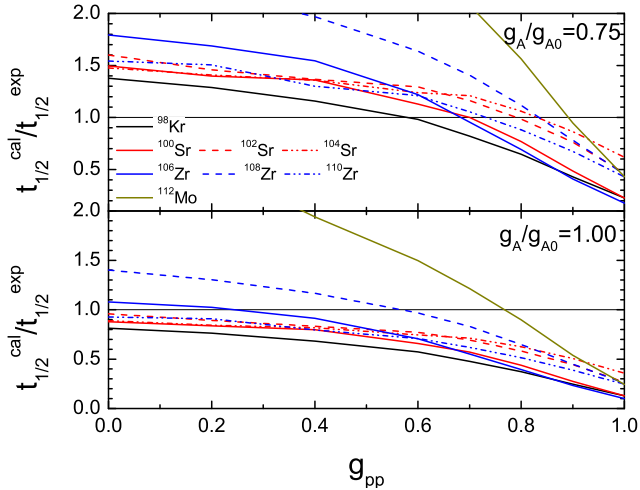}
\caption{The half-life dependence on $g_{pp}$ for different even-even isotopes.
The upper panel are results with quenched axial vector coupling constant
$g_A=0.75 g_{A0}$, while for the lower panel the bare one $g_{A0}=1.26$.}
\label{tpp}
\end{figure}

\begin{figure*}
\includegraphics[scale=0.9]{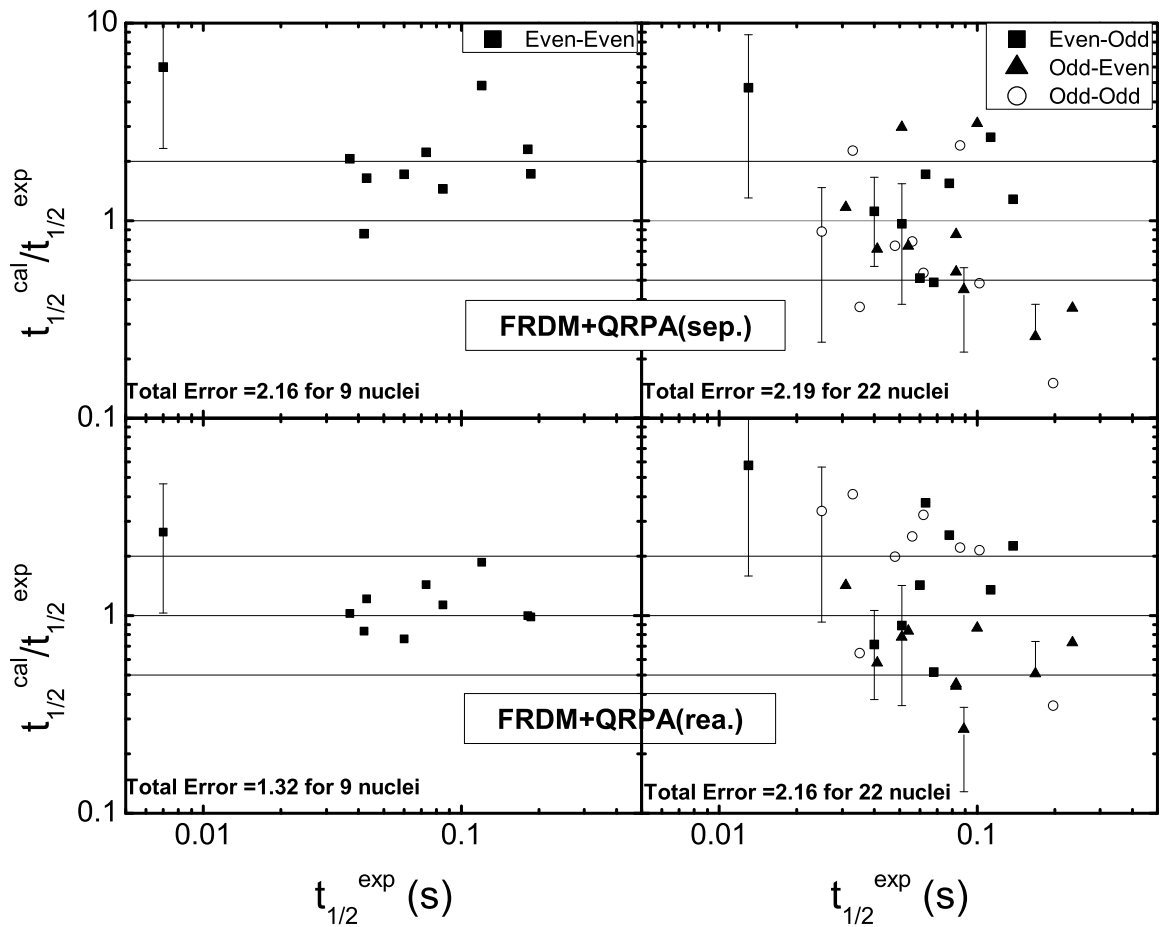}
\caption{A comparison of the calculated half-lives from Ref. \cite{MPK03} (upper
panels) and this work (lower panels) with the experimental ones from
RIKEN \cite{RIKEN11}. The experimental errors are taken into account for several
isotopes with large error bars.
The total errors are defined in \eqref{err}. The left panels are comparison for
even-even isotopes, and the right panels are for other types denoted by 
different symbols for the three cases of (Z, N).
The three horizontal lines in each panel correspond
to the ratios of 2, 1 and 0.5 respectively. Here, sep. is the abbreviation for
separable force and rea. for realistic force.}
\label{tcmp}
\end{figure*}

\section{results and discussion}

The single-particle wave functions and energy levels are obtained by solving
Schr\"odinger equation with the deformed Woods-Saxon potentials. The parameters
of the Woods-Saxon potential are taken from Ref. \cite{NBF86}. For the choice of
the
model space, based on previous experience, we start from the $0\hbar\omega$ to
one major shell above the Fermi surface of either proton or neutron, depending
on which is close to zero-energy, so for most isotopes here with $N\le 70$, the
model space is $0-5\hbar\omega$. For the deformation parameters, we choose
either the experimental ones if available (using the treatment for deformation
in Ref. \cite{FFRS11}) or those predicted by Ref. \cite{GCP09}. For the pairing
interaction, we adopt the Br\"uckner G-matrix with the Charge-dependent-Bonn
force. The detailed calculation procedure and choice of parameters are described
in Refs. \cite{RFSV07, YRFS08, FFRS11}. The same interaction is used as the
residual
interaction for pn-QRPA phonons.

As mentioned in the introduction, one of the new aspects of our work is
the introduction of a microscopic calculation for the first-forbidden decays.
Fig. \ref{ff} shows the region of nuclei covered in this paper.
For each nucleus we show the branching ratio for the forbidden
decay. They vary from several percent to at most $30\%$ in this region.
Thus our discussions below apply mainly to the Gamow-Teller decay
aspects of calculations.
In other regions, the FF contributions are more important and an accurate
determination of these decay widths can give a better accuracy, an example is
given in \cite{SYKO11} for $N=126$ isotones
where, for some isotones, the first forbidden decays contribute more than $80\%$
of the decay width. Overall, an
explicit calculation of FF decay is required for a complete account of
the beta-decay in the r-process path.

Two parameters are introduced as described in
Refs. \cite{YRFS08, FFRS11}: the renormalized particle-hole ($g_{ph}$) and
particle-particle ($g_{pp}$) strengths. However, the fitting procedure of these
parameters is a bit different from Refs. \cite{YRFS08,FFRS11}. For the
renormalized
particle-hole strength $g_{ph}$, the usual way is to fit the position of
the Gamow-Teller resonance. Since we don't have enough data in the Kr-Tc region of
interest, we adopt the same value as that derived in Ref. \cite{YRFS08} for
double-beta decay emitter $^{76}$Ge (In fact, the $\beta$-decay depends on the
low-lying
strength distributions and the choice of $g_{ph}$ does not affect the final
results too much).

The calculated half-lives are sensitive to the renormalized particle-particle
strength
$g_{pp}$. But we must also consider the possibility of
quenching of the axial-vector coupling constant
due to short-range correlations and to
multi-phonon effects which are excluded out in QRPA
calculations. Due to the lack of experimental data on log$ft$ values of single
decay branches in our mass region of interest, we
take the one used in Ref. \cite{FFRS11} that was derived from experiment
\cite{Guess11}
for
$^{150}$Nd, $g_A=0.75 g_{A0}=0.95$, 
where $g_{A0}=1.26$ is the bare value in the vacuum.

The relation between the calculated half-lives and values of $g_{pp}$ is
illustrated in Fig. \ref{tpp}. When $g_{pp}$ is increased, we obtain an enhanced GT
strength
to low-lying states, and therefore smaller calculated half-lives. From
Fig. \ref{tpp}, we find that without quenching, the half-lives are
underestimated, and the fitted $g_{pp}$ values are around zero. If
the quenching is included, realistic values from $0.6-0.9$ which reproduce the
half-lives of the isotopes are obtained, which agree well with the fitted $g_{pp}$
values of $\beta\beta$-decay half-lives in Ref. \cite{FFRS11}. 
Due to the large uncertainty in half-life of $^{100}Kr$ we exclude this
isotope from Fig. 1. Another isotope which is not included in the figure is
$^{114}$Mo, because it requires a larger model space due to its neutron number
$72$ (one more major shell should be added in the calculation compared with
other isotopes), hence a much longer time is needed for calculation of the whole range of
$g_{pp}$. However, as we shall see later, the results for $^{114}$Mo agree
well with those obtained with the $g_{pp}$=0.75 value we choose from the fitting.

\begin{figure}
\includegraphics[scale=0.45]{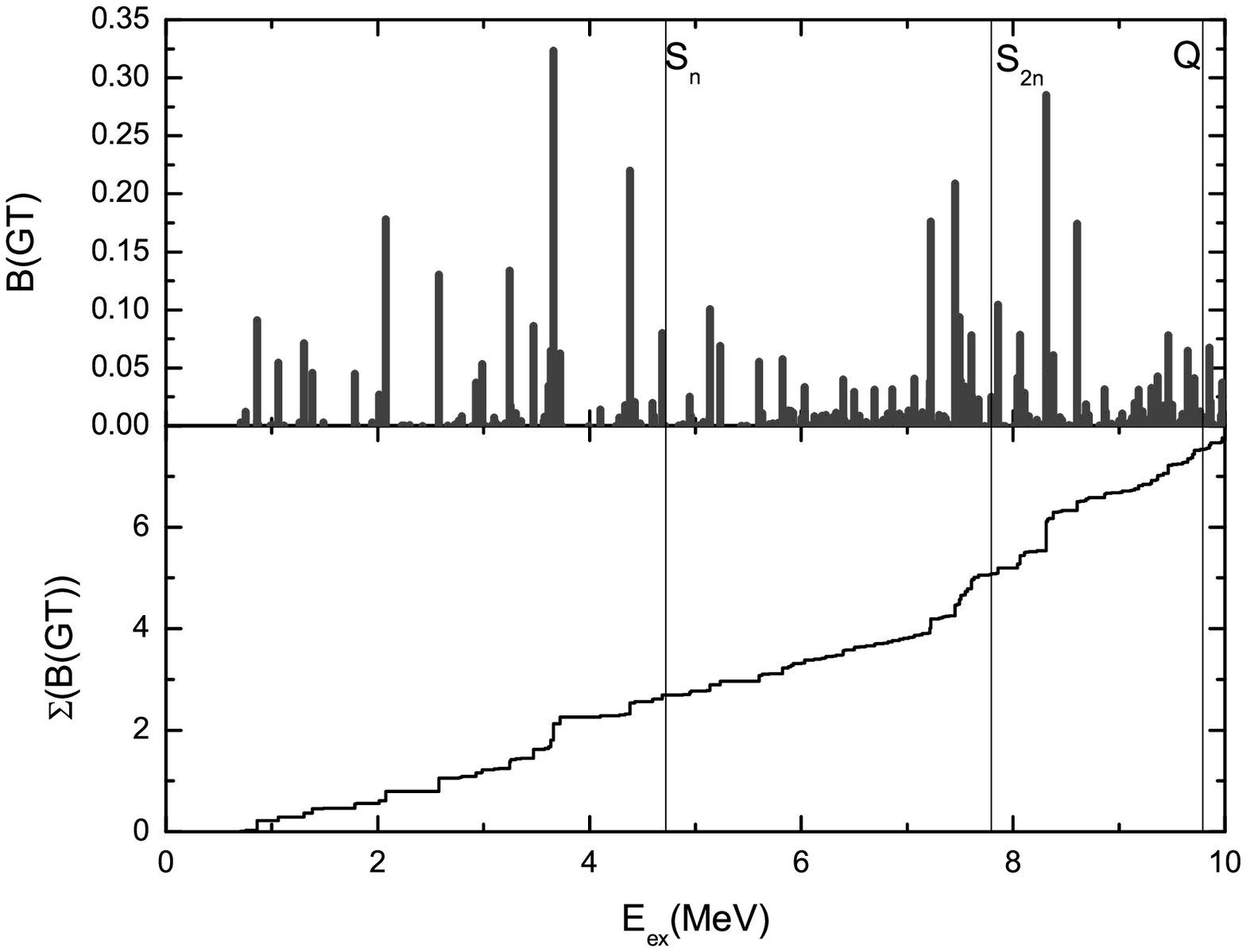}
\caption{As an example, the low-lying GT strength distribution (upper panel) and
GT running sum (lower panel) for $^{110}$Zr are shown.
The three vertical lines correspond to $S_n=4.72$ MeV, $S_{2n}=7.79$ MeV and
$Q=9.79$ MeV respectively, which are taken from Ref. \cite{Au02}. Here,
we
use $g_{pp}=0.75$, which gives a half-life $t_{1/2}=37.9$ ms and $\beta$-delayed
one-neutron emission probability $P_n=10.5\%$.}
\label{gtdis}
\end{figure}

\begin{figure*}
\includegraphics[scale=0.9]{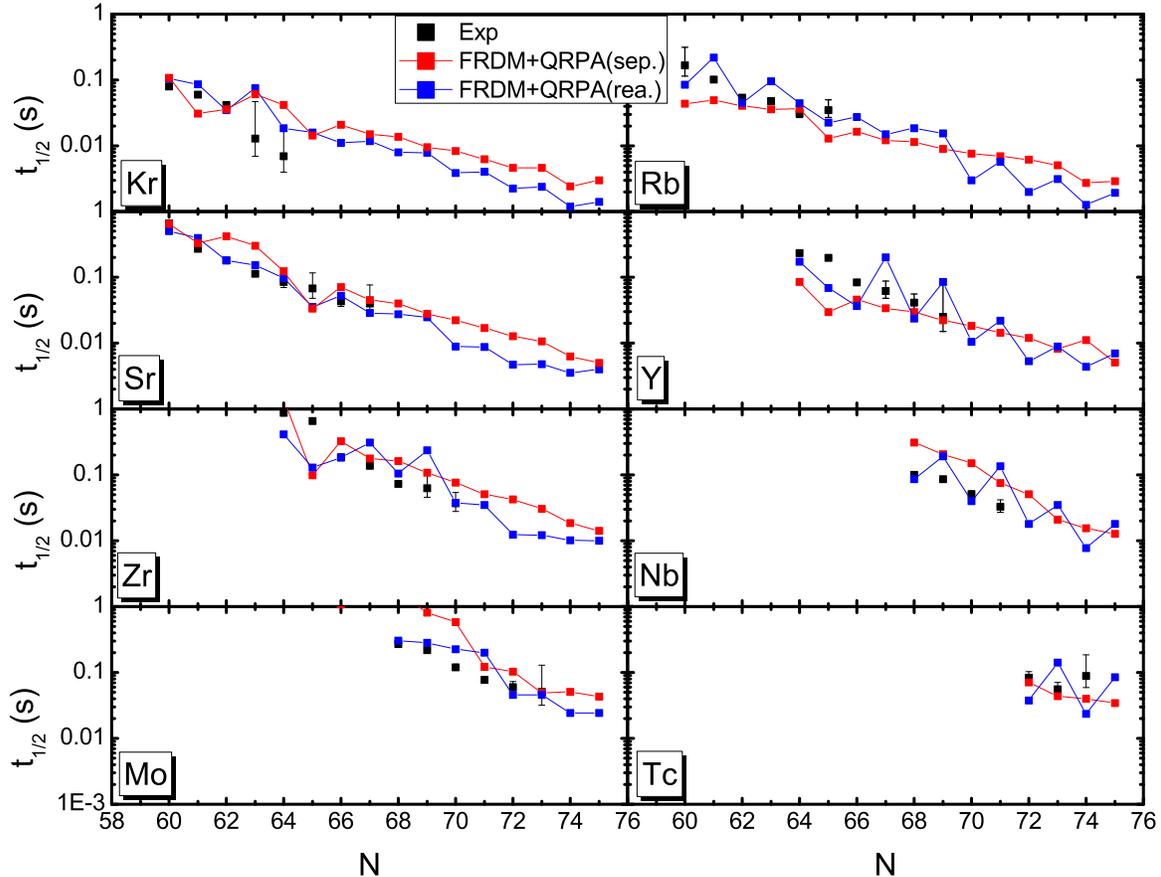}
\caption{A comparison among the calculated half-lives from Ref. \cite{MPK03}
(red),
this work (blue) and measured ones (if available) with error bars from
RIKEN \cite{RIKEN11} for Kr to Tc isotopes. The total errors are defined in
Ref. \cite{MPK03}. Here, as before, sep. is the abbreviation for separable force
and rea. for
realistic forces.}
\label{tcmpis}
\end{figure*}

\begin{figure*}
\includegraphics[scale=0.9]{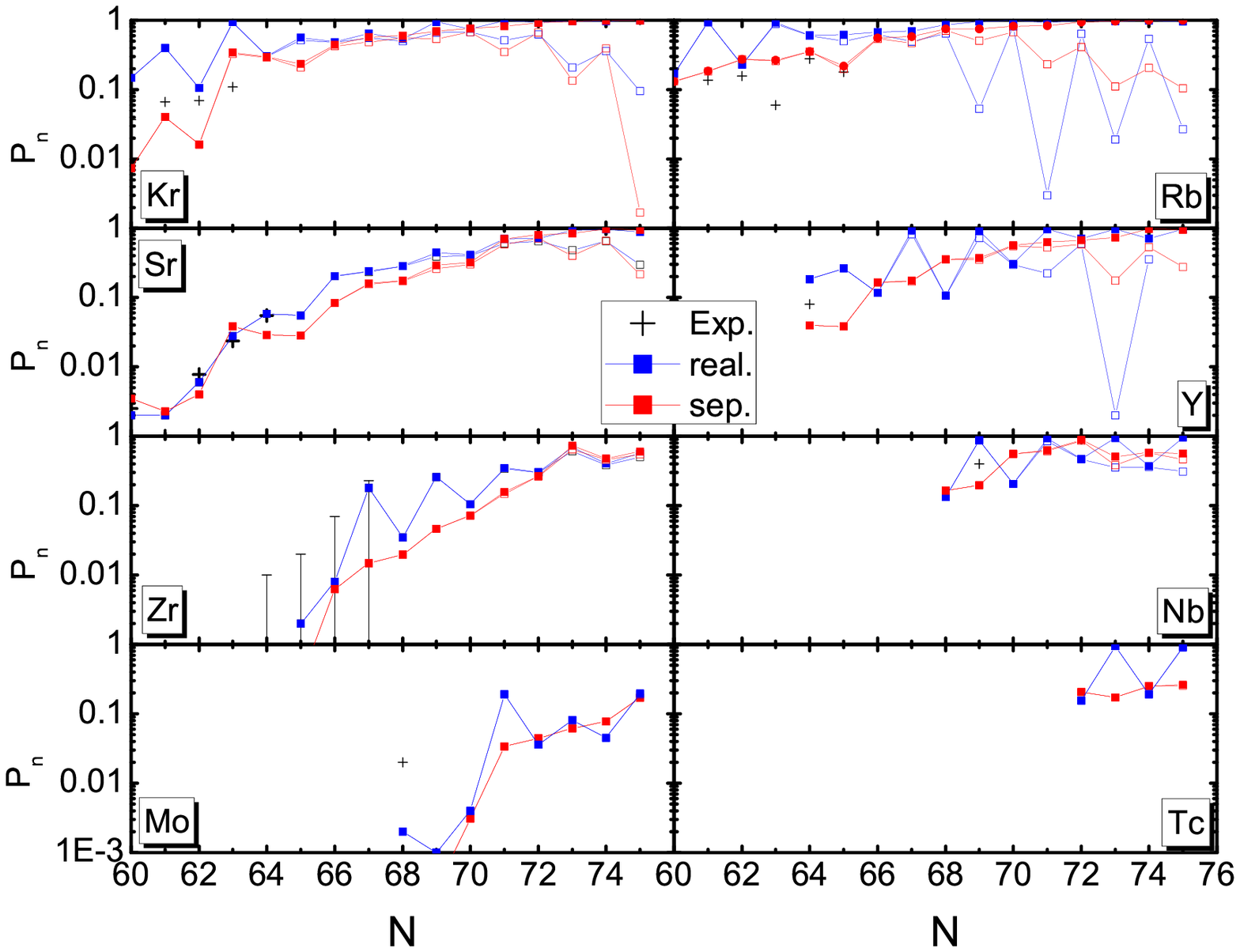}
\caption{A comparison among the calculated $\beta$-delayed neutron emission
probability $P_n$ values from Ref. \cite{MPK03} (red), this work (blue) and
measured
ones (if available) with error bars from RIKEN \cite{RIKEN11} for Kr to Tc
isotopes. Here sep. is the abbreviation for separable force and rea. for
realistic
forces. The empty squares are the $\beta$-delayed one neutron emission
probability $P_n$, while the solid ones are the total probability up to 3
neutron emission. }
\label{pn}
\end{figure*}

With the uncertainties of the choice of $g_{pp}$ from
$0.6-0.9$, the errors of the half-lives vary by a factor of two in general.
The optimal choice is $g_{pp}$=$0.75$ from the trends in the upper panel of
Fig. \ref{tpp}.
In Fig. \ref{tcmp}, we show the ratio between calculated and measured
half-lives with $g_{pp}=0.75$.
Following the definition from \eqref{err},
we obtain a total error of $2.1$
compared to $2.3$ in Ref. \cite{MPK03} for all nuclei from Kr to Tc.
If we compare our results with
those obtained in Ref. \cite{MPK03} (upper panels in Fig. \ref{tcmp}), we find
that we
have a better agreement for even-even nuclei (a total error of $1.32$ compared
to $2.16$ for 9 nuclei with reasonably small error bars) not simply because we
have
adjusted the parameters for this region, but mainly due to the adaption of
excitation energies relative to the ground states of the final odd-odd nuclei.
This gives a more accurate phase-space factors which effectively reduce
the half-lives, and gives improved agreement with the experiment. While for
Ref.
\cite{MPK03}, even without the effect of quenching, there is an over-estimation
for almost all even-even nuclei due to the excitation energies they choose.

As an example, we show in Fig. \ref{gtdis} the
low-lying GT strength distributions for $^{110}$Zr, in order to give a general
idea of how the low-lying strength is distributed, and how they contribute to
the decay width in deformed nuclei. Due to the deformations, the
contributions are split for $K=0$ and $K=\pm1$ parts, not only their energies
but also the strengths, this makes the distribution spread out. In
Fig. \ref{gtdis}, we show the position for Q value and the neutron separation
energies. The strengths with the lowest excitation energies are  most
important to the decay width because of their larger phase spaces. Thus,
nearly comparable amounts of GT strengths are located in the interval of
$S_n$-$S_{2n}$ and $S_{2n}$-$Q$, but the $\beta$-delayed two neutron emission
probability $P_{2n}$ is small compared with $P_n$. 
In this sense, one needs
both accurate predictions for the strengths and their positions, and the
advantage
of realistic force is that it provides a better determination of the excitation
energies. The effect of increasing $g_{pp}$ is that it enhances the
low-lying strength and shifts down the excitation energies, hence reduces the
half-lives.  
From the definition of $P_n$ in Refs. \cite{MPK03,MR90}, more
low-lying
strength below the neutron separation energies gives much smaller $P_n$ values,
and {\it vice versa}. Thus, a comparison with the experiments for both the
half-lives and the $P_n$ values is a good measure of how good the
nuclear structure descriptions are.

The agreement between experiment and theory for even-odd and odd-even isotopes
is about a factor of two worse than that for even-even isotopes.
The lack of particle-vibration coupling in the odd-mass
systems at this region does not seem to have too much affect on the
final half-lives, however. This is
consistent with the calculations in Ref. \cite{MPK03}, where a weak-coupling
approximation was assumed.

The agreement between experiment and theory is worse 
for the odd-odd isotopes, and there exists a 
systematic overestimation for the half-lives. This is
due to a shortcoming of the method we use, the lack of consideration of the
collectivity for even-even daughter nuclei. This over-predicts the energies
of the final states, and hence the calculated phase factors are smaller than
expected. However, in spite of the shortage of the methods, we can still keep the error
within an order of magnitude and for most isotopes approximately a factor of
five.
We note from \cite{RS12} that the r-process path does not depend strongly on the
beta
decay properties of the odd-odd nuclei due to their larger $S_n$ values.
Thus, instead of  
improving the models for odd-odd nuclei, one alternative way is to
simply use the average of the results for the neighboring odd-even
and even-odd nuclei for their values in an r-process database.

With the above comparisons and discussions, we extend our calculations to all
of the deformed Kr-Tc isotopes in the region $N=50-82$. We make
comparisons with experimental measurement (if available) and previous
theoretical predictions from Ref. \cite{MPK03}. The results are shown in
Fig. \ref{tcmpis}. The same set of Q-values taken from FRDM model as used in
Ref. \cite{MPK03} is adopted for the sake of comparison. One of the
differences
between our results and that of Ref. \cite{MPK03} is that the latter have
added an extra strength
spreading for each of the final states. In our calculations of even-even nuclei
the strength is already spread by the 
deformation effects, and there is not as much
motivation
for adding more by hand. But for our calculations of odd-mass and odd-odd
nuclei where the collective behavior has been excluded, the transitions
are just between the single-particle states
with little spreading compared to that in Fig. \ref{gtdis} for even-even nuclei.
In this
case, we might get better results by adding some spreading in the strength.

For most even-even isotopes, shorter half-lives by up to a factor of two are
predicted in our calculation compared to Ref. \cite{MPK03} due to the lower
excitation energies for the final states.
This behavior applies also to
some odd-mass nuclei.
For even-Z isotopes, the half-lives decrease with
the increase of neutrons with 
some small staggering behavior,
but overall agreement with
experiment is obtained. For odd-Z, there are systematic over estimations
of the half-lives for the odd-odd
isotopes with the reasons stated in Sec. II. 
In Ref. \cite{MPK03}, due to the additional strength spreading
put in by hand, they obtained a better agreements for these isotopes.
As discussed in Sec. II, the r-process results are not sensitive to
the half-lives of the odd-odd nuclei, and from a practical point of
view it would be adequate to simply use the average of the
calculated half-lives of the neighboring even nuclei for these odd-odd nuclei.

Another observable from the experiments for some isotopes is the
$P_n$ value, which with accurate separation energies gives a measure
of the Gamow-Teller strength distribution as shown in Fig. \ref{gtdis}.
From Fig. \ref{pn}, we find a good agreement again for even-even
isotopes from limited data, proving that the reliability of our descriptions for
deformed even-even isotopes in this region.
In Fig. \ref{pn}, one finds a staggering behaviours in the realistic
results for the even and odd $N$ number
neutrons especially for the odd-Z isotopes. The reason can be traced back to the
treatment of the odd nuclei with the lack of the collectivity. This shifts the
excitation energies up and the strength distributions are shifted systematically
to higher energies. This behavior is more obvious for odd-odd nuclei where nearly
all the strengths are shifted up. In applications of our calculations
to the r-process it is preferable 
to replace the calculated $P_n$ results for odd $N$ values with
the average of the neighboring even $N$ values.

\section{implications for the r-process}

 We have investigated how the r-process element abundances are affected by
the $\beta$-decay half-lives of various nuclei by changing the half-lives of
Moeller's predictions\cite{MPK03} for all even-even or odd-odd nuclei by one
order of magnitude.
These preliminary results agrees with recent r-process
simulations \cite{RS12}: for even-even nuclei, such changes of lifetimes have
tremendous effect on the peak formations, totally change the patterns of the
abundance distributions. 
In case of odd-odd nuclei, one order of magnitude change in all of the
half-lives results in essentially the
same the r-process abundance pattern except for the A=150-200 
mass region where the odd-even oscillation for elemental abundances 
are relatively changed by about a factor of two.
 
We can conclude from this simple simulation that current accuracy of 
deformed QRPA method can meet the needs of nuclear inputs for the
r-process simulation.
Our next step is to extend the
present calculations to other deformed regions, for example, the heavily deformed
rare-earth elements region, where the beta-decay data is limited,
and where it is still not understood how the the peak of rare-earth 
elements is formed. 
The final goal
is to calculate the $\beta$-decay properties over the whole nuclear chart. 
It is also
important to have reliable calculations for spherical nuclei, especially
those around $N=82$ that are important for the abundance peak around $A=130$.

\section{conclusion}
In this work, we investigated the $\beta$-decay properties of the 
Kr-Tc isotopes recently measured at RIKEN. 
With the pn-QRPA taking into account of realistic forces,
a good agreement has been obtained between the theory and the experiments
especially for even-even nuclei, with an accuracy within a
factor of two for most of them. 
The current calculations provide improved
results for the beta-decay half-lives of even-even nuclei. 
We plan to apply the present method to the rare-earth
region of deformed nuclei. We also plan to use the realistic
interactions for QRPA calculations of spherical nuclei.
This will eventually provide an improved set of predictions for
the half-lives and $P_n$ values that can be used  in r-process
network calculations for the element abundances.

This work was supported by the US NSF [PHY-0822648 and PHY-1068217]

\end{document}